\documentclass[aps,prl,twocolumn,showpacs,byrevtex]{revtex4-1}
\let\oldhat\hat
\renewcommand{\hat}[1]{\oldhat{\mathbf{#1}}}

\usepackage{times,mathptm}
\usepackage[dvips]{graphicx,color}

\begin{document}
\title{Stacking-sequence-independent band structure and shear exfoliation of two-dimensional electride materials}
\author{Seho Yi$^1$, Jin-Ho Choi$^{1,2}$, Kimoon Lee$^3$, Sung Wng Kim$^4$, Chul Hong Park$^5$, and Jun-Hyung Cho$^{1*}$}
\affiliation{$^1$ Department of Physics and Research Institute for Natural Sciences, Hanyang University, Seoul 133-791, Korea\\
$^2$ Research Institute of Mechanical Technology, Pusan National University, Pusan 609-735, Korea\\
$^3$ Department of Physics, Kunsan National University, Gunsan 573-701, Korea\\
$^4$ Department of Energy Science, Sungkyunkwan University, Suwon 440-746, Korea\\
$^5$ Department of Physics Education, Pusan National University, Pusan 609-735, Korea
}
\date{\today}

\begin{abstract}
The electronic band structure of crystals is generally influenced by the periodic arrangement of their constituent atoms. Specifically, the emerging two-dimensional (2D) layered structures have shown different band structures with respect to their stacking configurations. Here, based on first-principles density-functional theory calculations, we demonstrate that the band structure of the recently synthesized 2D Ca$_2$N electride changes little for the stacking sequence as well as the lateral interlayer shift. This intriguing invariance of band structure with respect to geometrical variations can be attributed to a complete screening of [Ca$_2$N]$^{+}$ cationic layers by anionic excess electrons delocalized between the cationic layers. The resulting weak interactions between 2D dressed cationic layers give rise to not only a shallow potential barrier for bilayer sliding but also an electron-doping facilitated shear exfoliation. Our findings open a route for exploration of the peculiar geometry-insensitive electronic properties in 2D electride materials, which will be useful for future thermally stable electronic applications.
\end{abstract}
\pacs{71.20.Ps, 73.22.-f, 81.07.-b}

\maketitle

\section{I. INTRODUCTION}

Since the successful exfoliation of graphene~\cite{Nov}, a two-dimensional (2D) hexagonal lattice of carbon atoms, in 2004, the search for new 2D materials has attracted a great deal of attention because of their promising prospects in both fundamental and applied research~\cite{Neto,Geim}. So far, a number of 2D materials have been discovered, including group-IV graphene analogs~\cite{Xu,Cah,Sah}, binary systems of group III-V elements~\cite{Sah,Nov2,Pac}, transition-metal dichalcogenides~\cite{TMDC1,TMDC2,TMDC3}, organometallic compounds~\cite{Wang,HJ}, and so on. Recently, it was successful to synthesize a layered electride material, dicalcium nitride Ca$_2$N (see Fig. 1), with a formula of [Ca$_2$N]$^{+}$$e$$^-$, where the anionic excess electrons are uniformly distributed in the 2D interstitial space between [Ca$_2$N]$^{+}$ cationic layers~\cite{Lee}. Motivated by such a pioneering work on the synthesis of Ca$_2$N, several candidates for 2D electrides have not only been proposed on the basis of density-functional theory (DFT) calculations, but also realized experimentally~\cite{Wal,Ino,Tada,Zhang,Ino2}.

Most of the developed 2D materials have been concentrated on the layered van der Waals (vdW) crystals~\cite{Butler}, where the layers are weakly bounded by vdW interactions. Although the vdW interactions in such 2D materials arise from the long-range, nonlocal electron correlations between interlayers, their band structures are known to vary sensitively with respect to the stacking sequence or the lateral interlayer shift. For instance, the band structure of bilayer graphene significantly changes even for a small lateral interlayer shift~\cite{son}. Meanwhile, the 2D electride materials can be classified as an ionic crystal with alternating cation and anionic layers~\cite{Lee}. Surprisingly, it is here demonstrated that the band structures of such layered-structure electride materials are nearly invariant with respect to the stacking sequence and the lateral interlayer shift. This peculiar electronic feature of 2D electrides which reflects a perfect isolated 2D electron system will provide an ideal platform for exploration of various exotic 2D phenomena such as charge density waves, spin ordering, and superconductivity~\cite{Gruner}.

\begin{figure}[ht]
\includegraphics[width=8.cm]{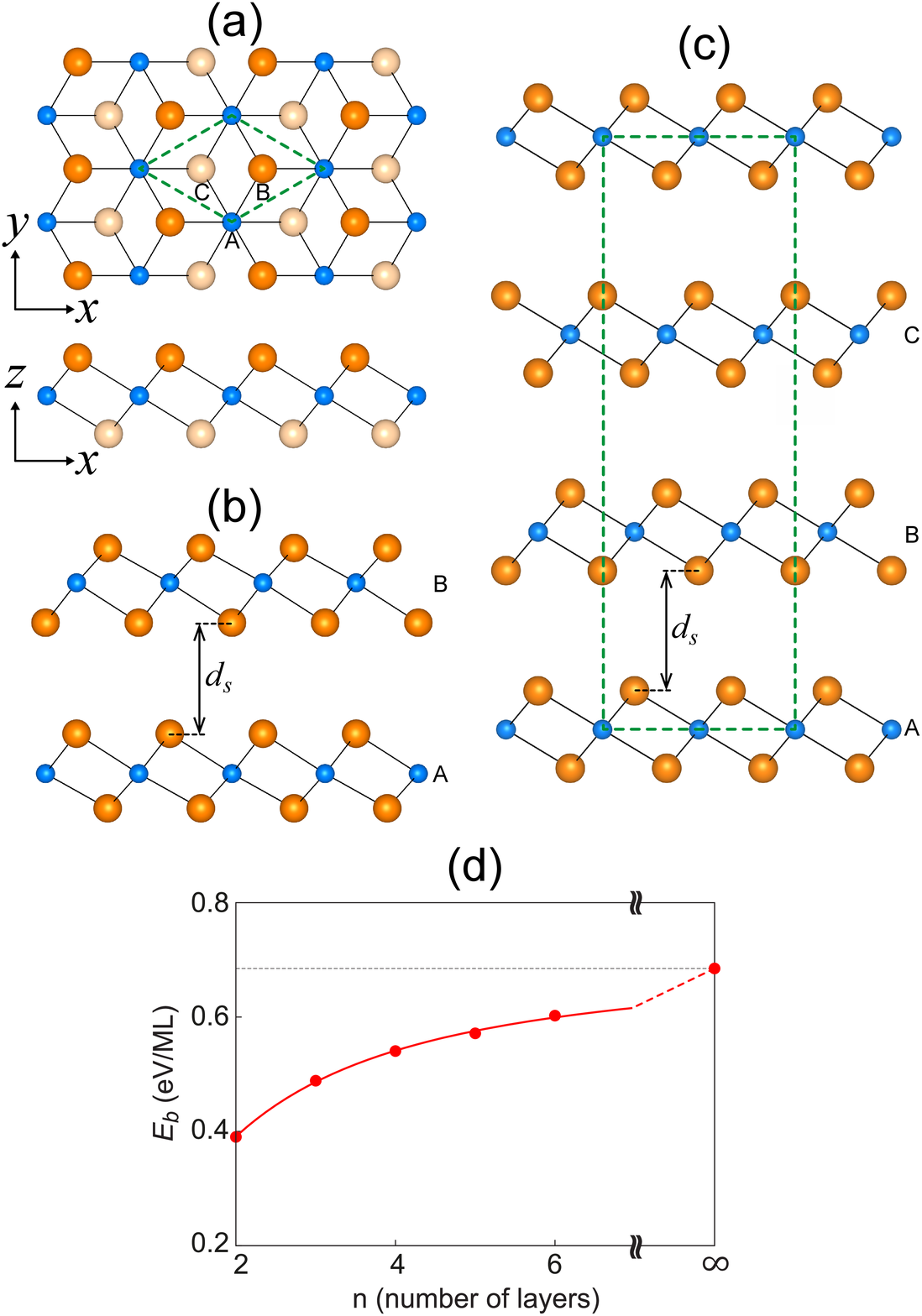}
\caption{(Color online) Optimized structures of the (a) ML, (b) bilayer, and (c) bulk Ca$_2$N. The large and small circles represent Ca and N atoms, respectively. In (a), the top (top panel) and side (bottom panel) views are given. For distinction, Ca atoms with different $z$ positions are drawn with dark and bright circles. In (b) and (c), the letters A, B, and C indicate the stacking sequences, representing the N-atom sites in the top view of (a). The dashed lines in (a) and (c) indicate the unit cells employed in the present calculation. The calculated binding energy as a function of the film thickness of Ca$_2$N is given in (d). The ${\infty}$ symbol in (d) represents the ABC bulk.}
\label{figure:1}
\end{figure}

In this paper, using first-principles DFT calculations, we systematically investigate how the electronic band structure of Ca$_2$N varies with respect to three different forms including monolayer (ML), bilayer, and bulk. We find that the band structure of either the bilayer or bulk changes little not only for its stacking sequence but also even for a lateral shift of layers. This invariance of band structure with respect to geometrical variations implies a complete screening of [Ca$_2$N]$^{+}$ cationic layers by the uniformly delocalized anionic excess electrons between cationic layers. Such a peculiar electronic feature in the 2D Ca$_2$N electride gives rise to not only a shallow potential barrier for bilayer sliding but also an electron-doping facilitated shear exfoliation. Interestingly, the band structure of the bilayer shows that the two bands originating from the N $p$ electrons in cationic layers and the anionic electrons between cationic layers touch at the ${\Gamma}$ point, forming a linear dispersion. However, this Dirac-like band dispersion is transformed into the usual parabolic dispersion in the bulk, caused by the periodic crystal potential along the direction normal to the [Ca$_2$N]$^{+}$ layers. The present findings further enrich our understanding of 2D electride materials for the design and development of future thermally stable electronic applications.

\section{II. CALCULATIONAL DETAILS}
Our DFT calculations were performed using the VASP code with the projector augmented-wave method~\cite{vasp1,vasp2}. For the treatment of the exchange-correlation energy, we employed the generalized-gradient approximation functional of Perdew,Burke, and Ernzerhof (PBE)~\cite{pbe}. The ML and bilayer were modeled by a periodic slab geometry with ${\sim}$30 {\AA} of vacuum in between the slabs. A plane wave basis was employed with a kinetic energy cutoff of 520 eV, and the ${\bf k}$-space integration was done with the 21${\times}$21 and 21${\times}$21${\times}$5 meshes in the Brillouin zones of the ML (or bilayer) and bulk, respectively. All atoms were allowed to relax along the calculated forces until all the residual force components were less than 0.01 eV/{\AA}.

\begin{table}[ht]
\caption{Calculated binding energy $E_b$ and interlayer separation $d_s$ [see Figs. 1(b) and 1(c)] of the bilayer and bulk of Ca$_2$N and Y$_2$C.}
\begin{ruledtabular}
\begin{tabular}{llcc}
  &  & $E_b$ (meV/ML) & $d_s$ ({\AA}) \\ \hline
Ca$_2$N  & AA bilayer  & 377 & 3.77 \\
  & AB bilayer  & 384 & 3.74 \\
  & AC bilayer  & 354 & 3.89 \\
  & ABC bulk    & 685 & 3.90 \\
  & ACB bulk    & 647 & 4.03 \\ \hline
Y$_2$C   & AA bilayer  & 523 & 3.40 \\
  & AB bilayer  & 552 & 3.34 \\
  & AC bilayer  & 484 & 3.56 \\
  & ABC bulk    & 1084 & 3.44 \\
  & ACB bulk    & 977 & 3.61 \\
\end{tabular}
\end{ruledtabular}
\end{table}

\section{III. RESULTS}

We begin to optimize the atomic structures of the ML; bilayer with stacking configurations AA, AB, and AC; and bulk with ABC and ACB [see Figs. 1(a), 1(b), and 1(c)]. For each optimized structure, we calculate the binding energy defined as $E_{\rm b}$ = ($nE_{\rm ML}$ $-$ $E_{\rm tot}$)/$n$, where $E_{\rm ML}$ is the total energy of the ML and $n$ is equal to 2 (3) for the bilayer (bulk). Figures 1(a), 1(b), and 1(c) show the most stable structures of the ML, bilayer, and bulk, respectively. We find that $E_{\rm b}$ increases as the film thickness of Ca$_2$N increases, converging to the value 685 meV/ML for the bulk [see Fig. 1(d)]. As shown in Table I, the AB bilayer is more stable than the AA and AC bilayers by ${\Delta}E_{\rm b}$ = 7 and 30 meV/ML, respectively. For the bulk, the ABC configuration is more stable than ACB by ${\Delta}E_{\rm b}$ = 38 meV/ML. The calculated interlayer separation [designated as $d_s$ in Figs. 1(b) and 1(c)] between neighboring [Ca$_2$N]$^{+}$ layers in the bilayer and bulk is also given in Table I. We find that $d_s$ is 3.74, 3.77, and 3.89 {\AA} for the AB, AA, and AC bilayers, while it is 3.90 and 4.03 {\AA} for the ABC and ACB bulk, respectively. These values for differently stacked bilayer or bulk indicate that $d_s$ tends to increase as $E_{\rm b}$ decreases. It is noted that, even though $E_{\rm b}$ of the bilayer is much smaller than that of the bulk (see Table I), $d_s$ is somewhat smaller in the bilayer compared to the bulk. Such relatively smaller values of $d_s$ in the bilayer can be associated with the fact that the anionic excess electrons are partially populated in the upper and lower regions of the bilayer (i.e., outside the interlayer region between the two [Ca$_2$N]$^{+}$ layers), as discussed below.

\begin{figure*}[h!t]
\includegraphics[width=18cm]{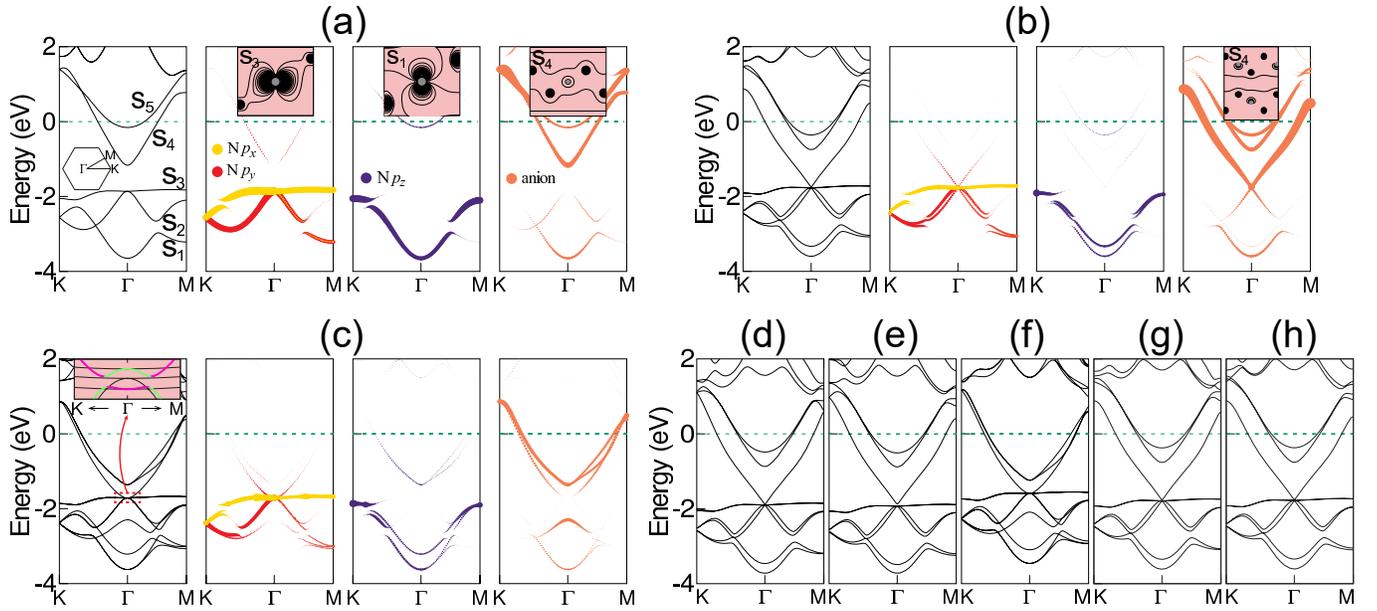}
\caption{(Color online) Calculated band structures of the (a) ML, (b) AB bilayer, and (c) ABC bulk. The bands projected onto the $p_x$, $p_y$, and $p_z$ orbitals of the N atom and the anionic excess-electron states in the interstitial region are displayed with circles whose sizes are proportional to the weights of such orbitals. The energy zero represents the Fermi level. The surface Brillouin zone is drawn in the inset of (a). A close-up of $S_2$, $S_3$, and $S_4$ states around the ${\Gamma}$ point is also displayed in (c). In the insets of (a) and (b), the charge-density contour plots of the $S_1$, $S_3$, and $S_4$ states at the ${\Gamma}$ point are drawn in the $xz$ plane containing N atoms, where the first line is at 0.3${\times}$10$^{-2}$ $e$/{\AA}$^3$ with spacing of 1.0${\times}$10$^{-2}$ $e$/{\AA}$^3$. The black and bright circles in the contour plots represent Ca and N atoms, respectively. For comparison, the band structures of the (d) AA bilayer, (e) AC bilayer, and (f) ACB bulk are given. The band structures of the laterally shifted AB bilayer configurations are also given in (g) and (h), where the upper layer in the AB bilayer is shifted by 0.5 {\AA} along the {\bf x} direction and 0.5 {\AA} along the {\bf y} direction, respectively.}
\label{figure:2}
\end{figure*}

Figures 2(a), 2(b), and 2(c) show the calculated band structures of the ML, AB bilayer, and ABC bulk, respectively, together with the band projection onto the N $p_x$, $p_y$, and $p_z$ orbitals and the anionic electron states distributed at the interstitial regions. Interestingly, we find that the band structures of the AA [Fig. 2(d)] and AC [Fig. 2(e)] bilayers are nearly the same as that of the AB bilayer. Also, the band structures of the ABC and ACB [Fig. 2(f)] bulks are very similar to each other. These results show that the band structure of either the bilayer or bulk is insensitive to its stacking sequence. To further confirm such a geometry-insensitive band structure in the AB bilayer, we laterally shift the two layers by 0.5 {\AA} along either the {\bf x} or {\bf y} direction, and find that even such lateral displacements change the band structure negligibly [see Figs. 2(g) and 2(h)]. This invariant electronic feature of Ca$_2$N is surprising in view of the fact that the band dispersion is usually sensitive to the crystal geometry. The presently predicted invariance of the cationic and anionic band dispersions with respect to the stacking sequence or the lateral interlayer shift implies that the [Ca$_2$N]$^{+}$ cationic layers are completely screened by the rather uniformly delocalized anionic electrons. By contrast, it was reported that the energy bands and the Fermi surface of bilayer graphene change significantly with respect to its lateral interlayer shift~\cite{son}. We note, however, that, despite the presence of the invariant band structure in Ca$_2$N, the stacking sequence influences to some extent the interlayer distance, which in turn affects the electrostatic Coulomb interactions between interlayers to yield the variation of $E_{\rm b}$ in the differently stacked bilayer or bulk (see Table I).

\begin{figure}[hb]
\includegraphics[width=8.5cm]{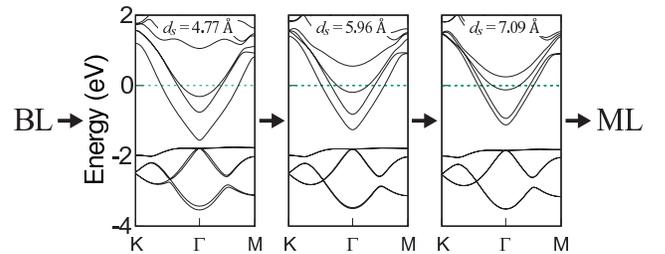}
\caption{Band structures of the AB bilayer (BL) with increasing $d_s$. The band structures of ML and BL are given in Fig. 2(a) and 2(b), respectively.}
\label{figure:3}
\end{figure}

\begin{figure}[ht]
\includegraphics[width=8.5cm]{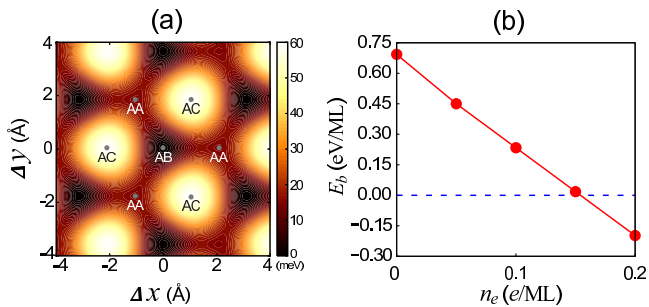}
\caption{(Color online) (a) Potential-energy surface for bilayer sliding, optimized by the lateral interlayer shift from the AB configuration. The potential energies of the AA, AB, and AC configurations are indicated by the points. (b) $E_b$ of the ABC bulk as a function of the additional excess electrons $n_e$.}
\label{figure:4}
\end{figure}

Recently, the angle-resolved photoemission spectroscopy experiment~\cite{ARPES} for bulk Ca$_2$N not only showed the existence of anionic states near the Fermi level $E_F$, but also measured the band dispersions and the Fermi-surface map, which agree well with the DFT results obtained using the PBE functional. It is, however, interesting to notice that our DFT calculations for the ML, bilayer, and bulk reveal a subtle variation of their band dispersions. As shown in Fig. 2(a), the band projection and charge characters of ML demonstrate that the $S_1$, $S_2$, and $S_3$ states are originated from the N $p_z$, $p_y$, and $p_x$ orbitals, respectively, while the $S_4$ and $S_5$ states with a parabolic dispersion along the $\overline{{\Gamma}K}$ and $\overline{{\Gamma}M}$ lines are mostly associated with the anionic excess-electron states in the upper and lower regions of the ML. It is noted that, although there is a sizable gap of 0.83 eV between $S_2$ (or $S_3$) and $S_4$ at the ${\Gamma}$ point, the $S_4$ and $S_5$ states have some hybridization with the $S_1$ and $S_2$ states [see Fig. 2(a)]. Meanwhile, the band structure of the bilayer shows the doublets of $S_1$, $S_2$, $S_3$, $S_4$, and $S_5$ [see Fig. 2(b)]. Interestingly, it is seen that the $S_2$ and $S_4$ states touch at the ${\Gamma}$ point, giving rise to a linear dispersion. The charge character of $S_4$ shows the anionic excess electrons residing in the interlayer region [see Fig. 2(b)]. In order to examine how the Dirac-like band dispersion changes with increasing $d_s$, we calculate the band structure of the bilayer as a function of $d_s$. As shown in Fig. 3, we find that the gap between $S_2$ and $S_4$ opens and increases as $d_s$ increases from the equilibrium interlayer distance of 3.74 {\AA}, thereby converging to the band dispersion of the ML. Noting that the differently stacked bilayers also have the Dirac-like band dispersion, it is likely that the ${\Gamma}$-point degeneracy of the $S_2$ and $S_4$ bands together with the nearly flat $S_3$ band is not related to symmetry but accidentally occurs with increasing their hybridization [see Fig. 2(b)]. However, this Dirac-like band dispersion is transformed into the usual parabolic dispersion in the bulk, which has the periodic crystal potential along the direction normal to the [Ca$_2$N]$^{+}$ layers. In the inset of Fig. 2(c), it is seen that the $S_2$ and $S_4$ bands around the ${\Gamma}$ point overlap and hybridize with each other.

It is interesting to examine how shear exfoliation is easily realized in the Ca$_2$N electride. For this, we calculate the potential-energy surface for bilayer sliding by optimizing the structure as one layer laterally shifts away from the other layer. The resulting contour plot is displayed in Fig. 4(a). We find that there is a minimum barrier of $E_s$ ${\approx}$ 17 meV for the sliding path from the AB to the AA bilayer. It is noteworthy that this value is comparable with that ($E_s$ ${\approx}$ 5 meV) for the case of sliding bilayer graphene with vdW interactions between layers~\cite{barrier}.
Such a marginal sliding barrier in the 2D Ca$_2$N electride is somewhat unexpected because conventional ionic compounds are usually formed from strong electrostatic interactions between ions. However, in Ca$_2$N electride material, the uniformly distributed anionic excess electrons in the 2D interstitial space between [Ca$_2$N]$^{+}$ cationic layers are likely to produce an effective screening of the cationic layers, thereby leading to an easy sliding in bilayer Ca$_2$N. Obviously, this underlying mechanism of easy sliding in 2D electrides contrasts with that in other 2D layered materials weakly bounded by vdW interactions. Meanwhile, since the $S_4$ and $S_5$ states near $E_F$ occupy the interlayer regions, it is natural to speculate that electron doping~\cite{doping} possibly increases $d_s$, therefore leading to a decrease in $E_b$. In order to estimate such charging effects on the interlayer interactions, we calculate $E_b$ while varying the magnitude of electron doping. Figure 4(b) shows the calculated values of $E_b$ for the ABC bulk as a function of the additional excess electrons $n_e$ ranging from 0 to 0.2$e$ per ML. We find that $E_b$ linearly decreases with increasing $n_e$ and finally becomes negative above $n_e$ ${\approx}$ 0.15 $e$ per ML [see Fig. 4(b)]. Therefore, we can say that electron doping can facilitate the shear exfoliation process in the 2D Ca$_2$N electride, leading to an easy fabrication of the ML, bilayer, or multilayer films.

\begin{figure}[ht]
\includegraphics[width=8.5cm]{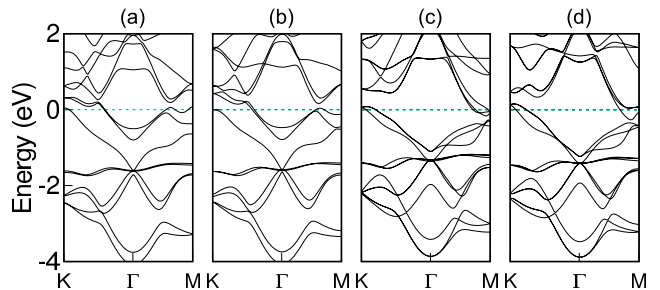}
\caption{Band structures of bilayer and bulk Y$_2$C: (a) AA- and (b) AB-stacked bilayers and (c) ABC- and (d) ACB-stacked bulks. The energy zero represents the Fermi level.}
\label{figure:5}
\end{figure}

In general, the electronic band structure of crystals is influenced by the periodic arrangement of their constituent atoms. In this sense, the prediction of geometry-insensitive band structure has not been seen in broad classes of 2D layered materials such as group-IV graphene analogues~\cite{Xu,Cah,Sah}, binary systems of group III-V elements~\cite{Sah,Nov2,Pac}, and transition-metal dichalcogenides~\cite{TMDC1,TMDC2,TMDC3}, and is realized only in 2D elecride materials where anionic excess electrons are confined within the interstitial regions between cationic layers. As another candidate of 2D electride materials, Y$_2$C was recently synthesized, which was described as [Y$_2$C]$^{2+}$2$e$$^-$ having two excess anionic electrons within the interlayer space~\cite{Zhang}. Our additional DFT calculations for bilayer and bulk Y$_2$C demonstrate that, although their values of $E_b$ and $d_s$ deviate depending on the stacking sequence (see Table I), their band structures change little with respect to the stacking sequence (see Fig. 5), indicating the feature of the geometry-insensitive band structure in 2D electride materials. In particular, the cationic and anionic layers of 2D electrides are two building blocks in their syntheses, which can influence the physical properties of 2D electronic systems. This aspect of 2D electrides will surely enrich the exploration of several exotic 2D electronic phases, including charge- and spin-density-wave condensates, magnetism, and superconductivity. The theoretical exploration of such a fascinating regime will be the subject of future work.

\section{IV. SUMMARY}
We have performed a comprehensive DFT study for the ML, bilayer, and bulk Ca$_2$N to investigate their electronic band structures while varying the stacking sequence and lateral interlayer shift. We found that such geometrical variations change the band structure of the bilayer or bulk, little due to a complete screening of [Ca$_2$N]$^{+}$ cationic layers by the surrounding anionic electrons. Such a peculiar electronic feature in the 2D Ca$_2$N electride gives rise to not only a shallow potential barrier for bilayer sliding but also an electron-doping facilitated shear exfoliation. Our findings open a route for designing 2D electride materials with geometry-insensitive electronic properties, which will be useful for future thermally stable electronic applications.
\vspace{0.4cm}

\section{ACKNOWLEDGEMENTS}
This work was supported by the National Research Foundation of Korea (NRF) grant funded by the Korean government (2015M3D1A1070639). The calculations were performed by KISTI supercomputing center through the strategic support program (KSC-2016-C3-0001) for the supercomputing application research.

\noindent $^{*}$ Corresponding author: chojh@hanyang.ac.kr


\end{document}